\documentclass{article}
\usepackage[utf8]{inputenc}

\usepackage{tabu}
\usepackage{multirow}
\usepackage{adjustbox}

\usepackage{forest}
\usepackage{tikz-qtree}
\usepackage{tikz}
\usetikzlibrary{shapes.geometric}
\usetikzlibrary{arrows.meta,arrows}

\graphicspath{{images/}} 

\usepackage{subfig}
\usepackage{multirow}

\usepackage{setspace} 

\title{\textbf{Modeling Ground-to-Air Path Loss for Millimeter Wave UAV Networks}}

\author{
	\textbf{Hazim~Shakhatreh$^{1}$\thanks{hazim.s@yu.edu.jo}~, Waed Malkawi$^{1}$,Ahmad Sawalmeh$^{2}$, }\\{ \textbf{Muhannad Almutiry$^{3}$,Ali Alenezi$^{3}$ }}\\
$^{1}$Department of Telecommunications Engineering, Hijjawi Faculty for Engineering \\ Technology, 
Yarmouk University,Irbid, Jordan,\\
$^{2}$Computer Science Department, Northern Border University, Arar, Saudi Arabia
\\
$^{3}$Electrical Engineering Department, Northern Border University, Arar, Saudi Arabia
}

\begin{document}
	\maketitle
	
	\begin{abstract}
		Path loss is a significant component of wireless communication channel design and analysis and reflects the reduction in a transmitted signal's power density. Due to the differences in the propagation conditions, wireless aerial channels' features differ from those of terrestrial wireless channels; therefore, unmanned aerial vehicle path loss models are often different from conventional terrestrial wireless channel path loss models. A mathematical propagation model is proposed in this paper to estimate the Ground-to-Air path loss between a wireless device and a low-altitude platform using the frequency bands of the millimeter wave. The suggested model of Ground-to-Air path loss will assist academic researchers in formulating several vital problems.\\

	\end{abstract}

	\onehalfspacing
	
	\section{Introduction}
	\label{Int}
	UAVs are growing exponentially in numerous civilian applications including wireless coverage, search and rescue, and real-time surveillance.~\cite{r1}. UAVs can be used for wireless coverage in emergency cases where each UAV is used as an aerial wireless base station when the terrestrial network goes out of operation~\cite{r2}. It can also be used to provide better coverage for users and higher data speeds~\cite{r3}. The authors in~\cite{r4_2} suggested a path loss model between ground wireless devices and a low-altitude platform used as an aerial wireless base station, which enabled academic researchers to formulate several significant problems. The path loss model showed a strong propensity for two distinct types of propagation, which for outdoor receivers were extensively studied and evaluated. Within this model the authors of~\cite{r5} defined the tradeoff; The path loss between the low altitude platform and the ground wireless device increases at a higher altitude, but also increases the possibility of a line of sight connections. On the other side, there are low probability line of sight links at low altitude, while the loss of the path decreases. However, this model assumed a downlink scenario, from low altitude platform to a terrestrial terminal, and the frequency bands are 5800,2000 and 700 MHz. 
	The applicability of this model is constrained by these assumptions when the uplink scenario from terrestrial terminal to the low altitude platform is considered, and when the millimeter wave frequency bands is used.
	
	The uplink scenario in which the data is transmitted to the UAV from the ground wireless devices is considered by just a few studies. Owing to the limited transmitting capacity of wireless devices, users can not be able to communicate in emergencies with remote undamaged ground stations (such as tsunamis, earthquakes or floods).
	Moreover, the wireless devices could not be able to recharged due to physical damage of the energy infrastructure. 
	In such cases, providing stranded people with wireless coverage becomes very critical as people in the disaster-affected area seek emergency information, locate family members and friends and get advice to help evacuate the area affected by the disaster~\cite{r6}. 
	Noting that in catastrophic circumstances, energy efficiency is more important for user devices that can be controlled by studying the uplink scenario than UAV energy efficiency because UAV can move long distances for the recharging process.
	The potential for combining millimeter wave communications with UAV networks is discussed in~\cite{r7} to meet the high throughput requirements of most UAV applications. Deploying mmWave communications with UAV networks has two major advantages. 1) The large range of mmWave is capable of greatly enhancing the capacity of UAV networks, thereby meeting the requirement for prompt responses.
	2) Through using mmWave, UAV network data traffic can be substantially increased, as in short-range transmissions, mmWave communications can deliver high throughput~\cite{r7}.
	
	In this research work, we suggest a mathematical propagation model using mmWave frequency bands to estimate a Ground-to-Air (GTA) path loss between a terrestrial terminal and a low altitude platform. Developing an RF-model requires a specific study description factors and constraints; the characteristics of the buildings are one of the most critical conditions in an urban setting.~\cite{r4_2}. 
	Within this prediction model four simulation environments are used: 1) Suburban Environment, 2) Urban Environment, 3) Dense Urban Environment, 4) Urban Environment with high-rise buildings. 
	The behavior of GTA channels for mmWave bands is investigated in this work at two separate frequencies: 28 GHz and 73 GHz, which are believed to cover a broad spectrum of applications.
	For each frequency band, the analysis and simulations are carried out over four different environments: suburban, urban, dense urban and high-rise buildings environments. Researchers will measure the predicted path loss for wireless devices on millimeter UAV networks with the proposed GTA path loss model. Numerous parameters, such as the SINR and the throughput, can be identified using path loss. 
	To the best of our knowledge, our research work is the first work that proposes a path loss model for  Ground-to-Air communication system at Millimeter Wave frequency bands in UAV Networks. 
	
	
	The rest of this paper is structured as follows. In Section \ref{related_work}, the related works is presented. Section \ref{GTA} discuss the Ground-to-Air path loss model for four simulation environments, namely, the so-called (1) Suburban Environment that includes rural areas, (2) The most popular Urban Environment, which represents the typical European city, then, (3) Dense Urban Environment that reflects those types of towns where buildings are near to each other, Next, (4) High-rise Urban Environment, reflecting new towns with the skyscrapers.  
	Finally, Section \ref{con} presents the conclusion and future work.
	
\section{Related Works}
\label{related_work}

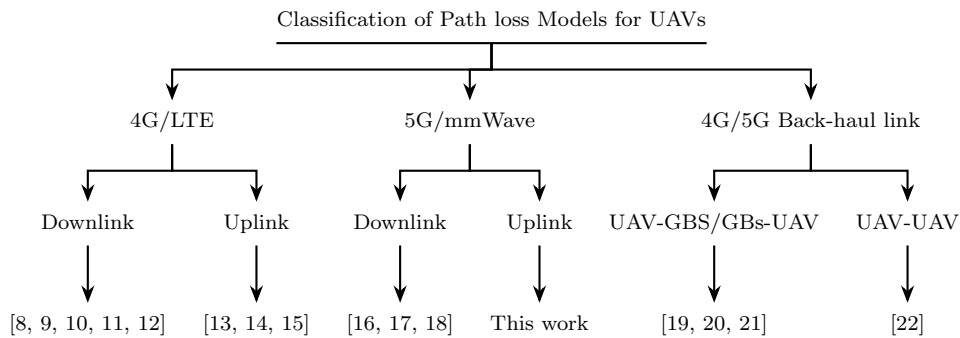
\begin{figure*}[h]
	\centering
	\begin{forest}
		for tree={
			align=center,
			parent anchor=south,
			child anchor=north,
			font=\footnotesize,
			edge={thick, -{Stealth[]}},
			l sep+=10pt,
			edge path={
				\noexpand\path [draw, \forestoption{edge}] (!u.parent anchor) -- +(0,-10pt) -| (.child anchor)\forestoption{edge label};
			},
			if level=0{
				inner xsep=0pt,
				tikz={\draw [thick] (.south east) -- (.south west);}
			}{}
		}
		[  Classification of Path loss Models for UAVs
		[4G/LTE
		[Downlink
		[
		\cite{feng2006path,al2014modeling,matolak2016air,sun2016air,matolak2017air}
		]
		]
		[Uplink
		[
		\cite{yang2018energy,zeng2016throughput,ranjan2018study}
		]
		]
		]
		[5G/mmWave
		[Downlink
		[
		\cite{khawaja2017uav,khawaja2018temporal,yang2019machine}
		]
		]
		[Uplink
		[
		This work
		]
		]
		]
		[4G/5G Back-haul link 
		[UAV-GBS/GBs-UAV
		[
		\cite{al2017modeling,shi2018multiple,gapeyenko2018flexible}
		]
		]
		[UAV-UAV
		[
		\cite{azari2019cellular}
		]
		]
		]
		]
	\end{forest}
	\caption{Classification of path loss models for UAVs Based on Operation Frequency Band.}
	\label{pathloass}
\end{figure*}

Path loss models play an important role in the designing and analyzing of wireless communication systems. It reflects the reduction amount in the power density of a transmitted signal.
Due to the variations in the propagation environments, including ground to air and air to ground communication links for UAVs networks, the UAVs network's path loss models are different from the conventional path loss models for cellular networks. UAVs network can be classified based on the operating frequencies,
Figure~\ref{pathloass} presents the classification of the path loss models for UAVs based on the operating frequency bands. This Figure shows three bands for path loss models used in UAVs communication, which are: (1) Fourth generation (4G) long term evolution (LTE) operating in microwave bands. (2) Fifth-generation (5G) operating in a mmWave bands. (3) The back-haul communication link between UAV and ground base station (GBS) can be worked either in 4G or 5G bands.

\subsection{4G/LTE Path loss Models}
Many studies in the literature proposed path loss models for UAV-based aerial communication frameworks in downlink and uplink scenarios for different environments, such as overseas, suburban, urban, dense urban, and high-rise building environments.

The authors in \cite{feng2006path}, developed a statistical propagation path loss model for air to ground (ATG) channels in an urban environment. In this model, a downlink scenario for ATG communication channel between an aerial base station and the mobile terminal is proposed; furthermore, it can be worked at frequencies between 200MHz - 5GHz. In this model, the radio channel is classified into three propagation groups a line of sight, obstructed line of sight, and a non-line of sight. The ATG model provides a higher line of sight and less non-line of sight probabilities compared to the terrestrial communication platform.

While the work in \cite{al2014modeling}, proposed an ATG path loss model for UAVs network, they developed a statistical propagation model for predicting the path loss between an airborne base station at a low altitude platform (LAP) and a ground terminal. The ray-tracing simulator was used to model three different types of rays, namely: 1)direct rays, 2)reflected rays, 3)diffracted rays for a line of sight and non-line of sight channel in different environments. The path loss model is found as a function of the aerial base station's altitude and the radius of the coverage area.

Moreover, in \cite{matolak2016air,sun2016air,matolak2017air}, Matolak et al. developed measurement-based path loss models for ATG channel for three different environments namely, Overwater \cite{matolak2016air}, Hilly and Mountainous \cite{sun2016air} and Sub-urban and Near-Urban \cite{matolak2017air} environments.  These works aim to provide a model that can be used in the evaluation of UAVs as aerial communication systems. They described ATG channel model in both dual-band (L-band with 970 MHz and C-band with 5 GHz) for over water, hilly and mountainous, suburban and near-urban environments that include propagation path loss,root-mean-square delay spread (RMS-DSs), small-scale RicianK-factor to develop a path loss model and wide-band tapped-delay lines (TDL) third-ray dispersive channel models.

On the other hand, the ground to air GTA wireless communication system between ground terminals (GT) and UAV is studied in \cite{yang2018energy,zeng2016throughput,ranjan2018study}. The authors in  \cite{yang2018energy}, proposed a ground to UAV (GTU) wireless communication system, where UAV acts as a UAV enabled data collection system, to collect data from GT. They found a power consumption trade-off between served GTs and the UAV in GTU uplink communication system considering the trajectories of UAV. In this system, the required transmission power of the GT to send their data can be reduced if the UAV flies close to GT to establish better GTU communication links. Therefore, the aim of this study is to characterize the trade-off for the GTU communication system to find the optimal transmission power of the GT and the best trajectory of the UAV. In \cite{zeng2016throughput}, the authors utilized a free space path loss to developed an uplink communication system between mobile relying node (UAV) and GTs. They studied the problem of throughput maximization to optimize the UAV trajectory and the GT-to-UAV power allocation. The authors in \cite{ranjan2018study}, analyzed different path loss models for downlink and uplink channel in emergency UAV-based wireless communication system. For the uplink communication scenario, they utilized  Winner II Uplink (WinIIU) \cite{meinila2009winner} and Two-ray uplink path loss models. 
	
\subsection{5G/mmWave Path loss Models}

Millimeter-wave (mmWave) frequency bands provide massive bandwidth that can be used in 5G networks to significantly increase the data rate. 
Many works in the literature were conducted to develop and investigate path loss models in downlink scenarios (ATG) for mmWave \cite{khawaja2017uav,khawaja2018temporal,yang2019machine}.

The authors in \cite{khawaja2017uav}, studied the large scale characterization of mmWave ATG channels for UAVs communication in the time domain. A ray-tracing simulation with a Wireless InSite simulator was conducted to study the behavior and characteristics of the ATG mmWave bands for frequency bands 28 and 60 GHz. They considered four environments, namely: overseas, rural, suburban and urban to study the received signal strength (RSS) and root mean square delay spread of multipath components at UAV altitudes between 20 to 150 meters.  The results showed that the RSS follows the two ray propagation model \cite{parsons2000mobile,matolak2016air} that adds the sea/earth surfaces reflection to the line of sight component.

On the other hand, the same authors of \cite{khawaja2017uav} analyzed in \cite{khawaja2018temporal}
a small scale characteristics of the mmWave propagation channels for UAVs communication in time and spatial domains for oversea, rural, suburban and dense-urban environments using Wireless InSite ray-tracing simulator using Omni-directional antennas for the receivers and transmitter. The results showed that the ATG mmWave propagation channel's received multipath components could be grouped into two components: persistent and non-persistent components, where the persistent component consists of the line of sight and ground reflected component, whilst the non-persistent contains the non-line of sight components. 

Moreover, the work in \cite{yang2019machine}, proposed a prediction method based on machine learning for path loss and delay spread in ATG mmwave channels. They employed two algorithms, which are random forest and K-nearest neighbors, for the prediction method. Then, they proposed a feature selection scheme of the proposed machine learning method for prediction accuracy improvement. A ray-tracing software was utilized in an urban environment at 28 GHz to generate the data that can be used for performance verification of the proposed prediction path loss model.   

\section{Ground-to-Air Path Loss Model}
\label{GTA}
\begin{figure*}[!h]
	\centering
	
	\subfloat[Top view of the targeted area -  Suburban.]{\label{C1-figur:1}\includegraphics[width=.33\textwidth]{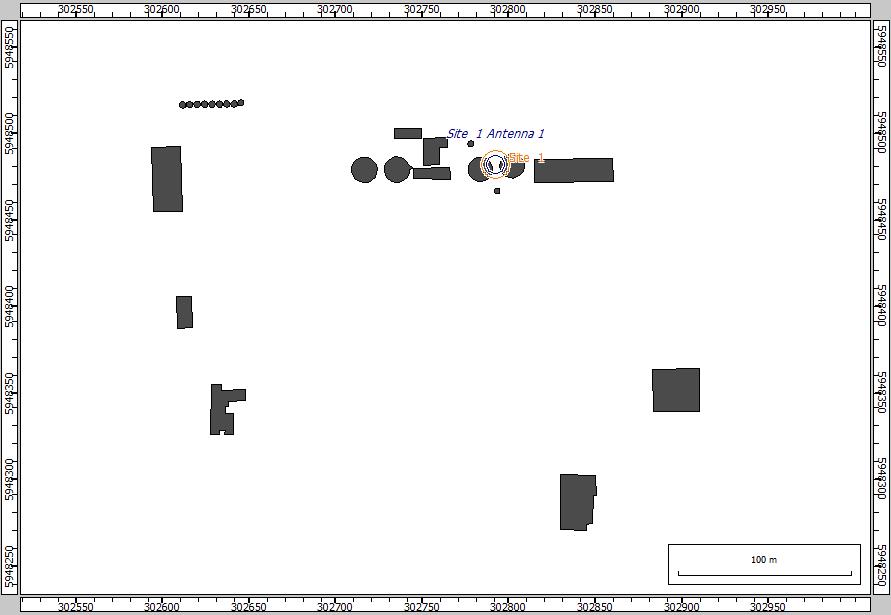}}
	\subfloat[3D view of the targeted area - Suburban.]{\label{C1-figur:2}\includegraphics[width=.33\textwidth]{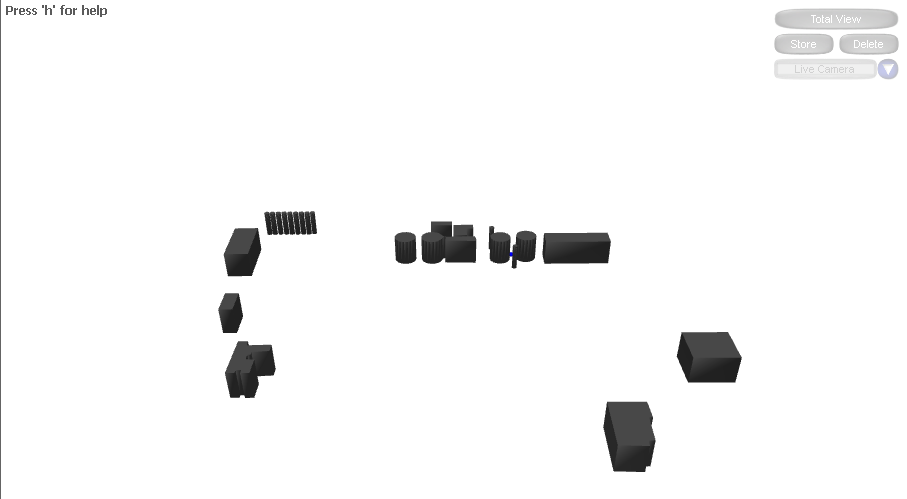}}
	\subfloat[Colored map of the path loss - Suburban.]{\label{C1-figur:3}\includegraphics[width=.33\textwidth]{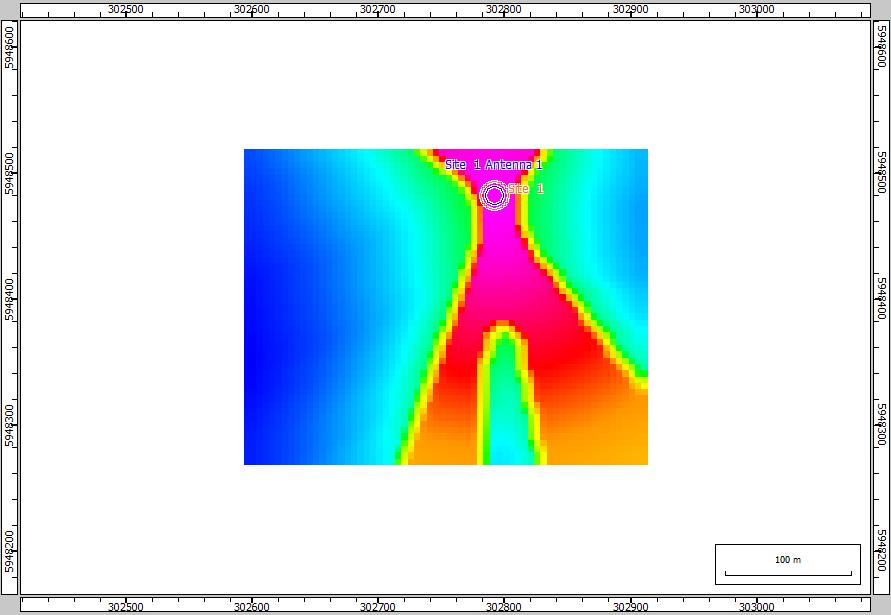}}
	\\
	\subfloat[Top view of the targeted area - Urban.]{\label{C2-figur:1}\includegraphics[width=.33\textwidth]{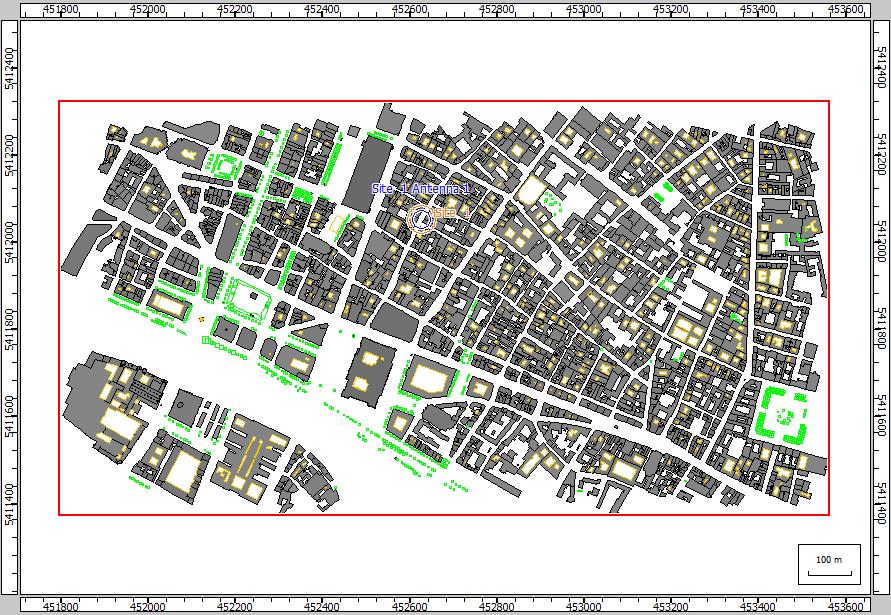}}
	\subfloat[3D view of the targeted area - Urban.]{\label{C2-figur:2}\includegraphics[width=.33\textwidth]{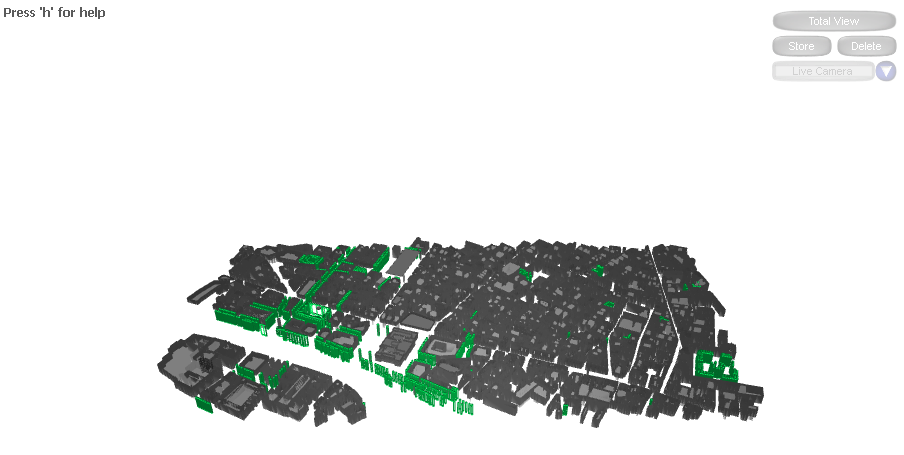}}
	\subfloat[Colored map of the path loss - Urban.]{\label{C2-figur:3}\includegraphics[width=.33\textwidth]{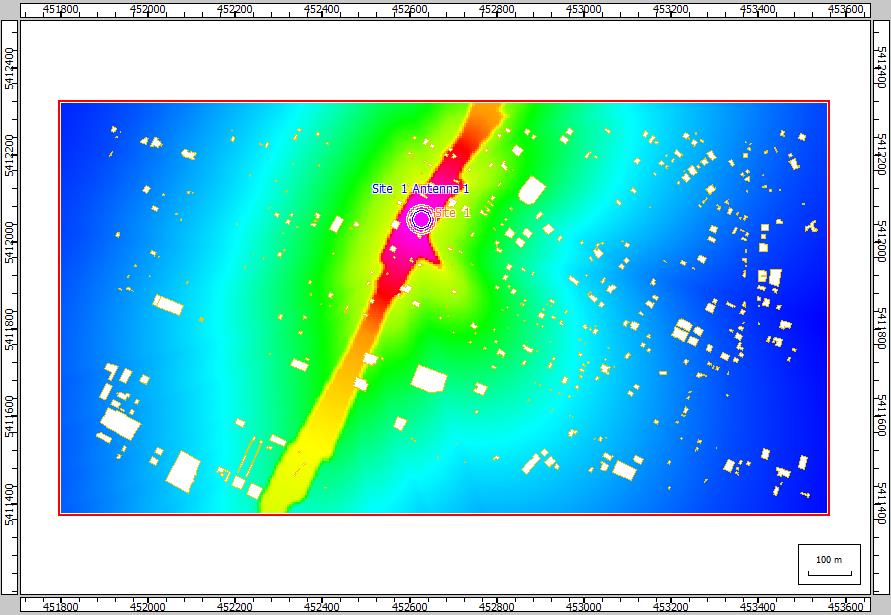}}
	\\
	\subfloat[Top view of the targeted Dense-Urban.]{\label{C3-figur:1}\includegraphics[width=.33\textwidth]{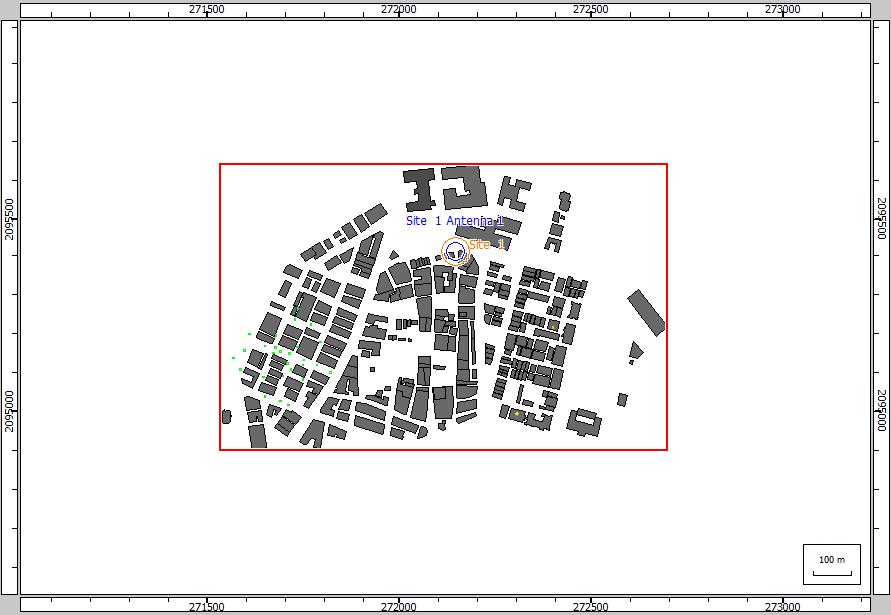}}
	\subfloat[3D view of the targeted Dense-Urban ]{\label{C3-figur:2}\includegraphics[width=.33\textwidth]{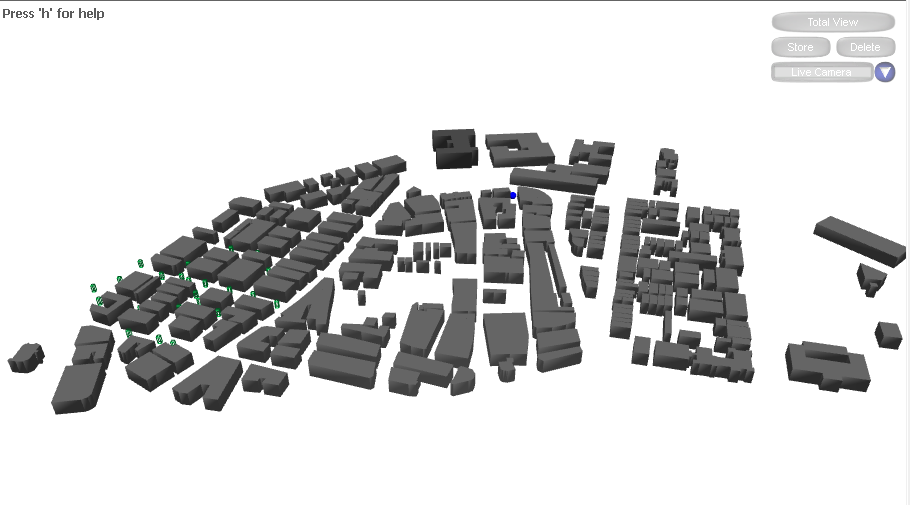}}
	\subfloat[Colored map of the path loss - Dense-Urban.]{\label{C3-figur:3}\includegraphics[width=.33\textwidth]{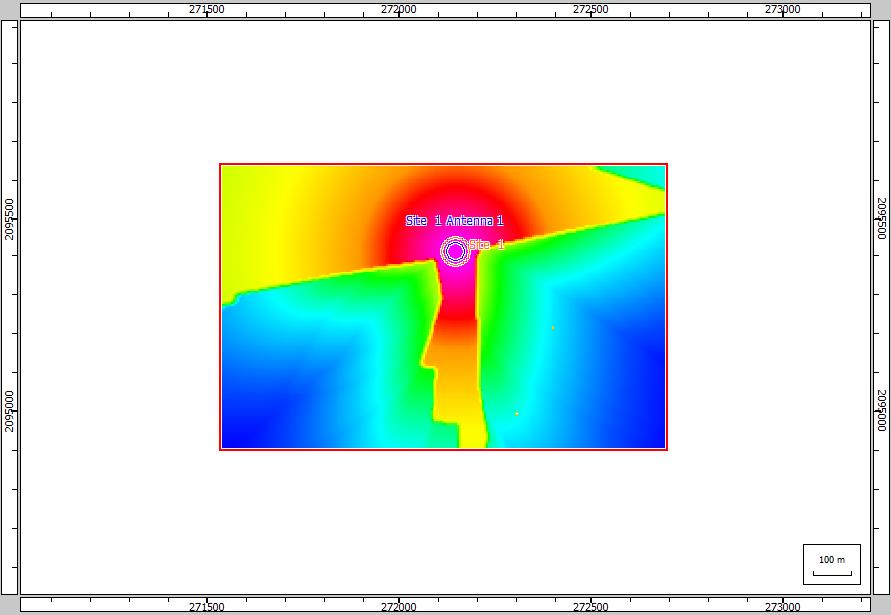}}
	\\
	\subfloat[Top view of the targeted area - High-rise building.]{\label{C4-figur:1}\includegraphics[width=.33\textwidth]{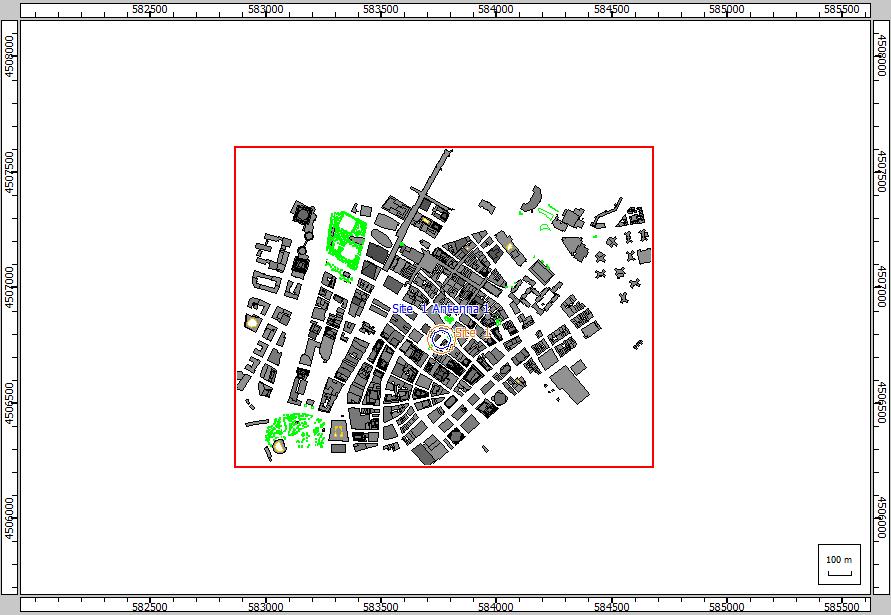}}
	\subfloat[3D view of the targeted area - High-rise building.]{\label{C4-figur:2}\includegraphics[width=.33\textwidth]{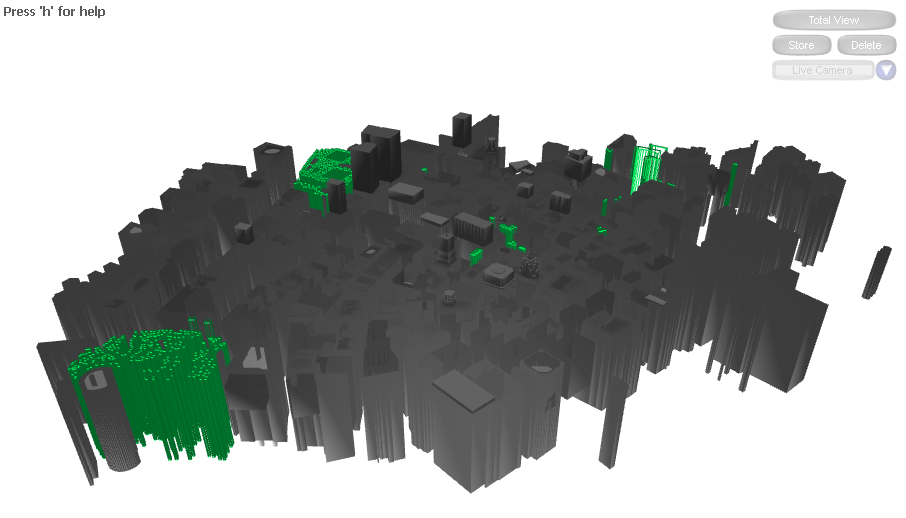}}
	\subfloat[Colored map of the path loss - High-rise building area. ]{\label{C4-figur:3}\includegraphics[width=.33\textwidth]{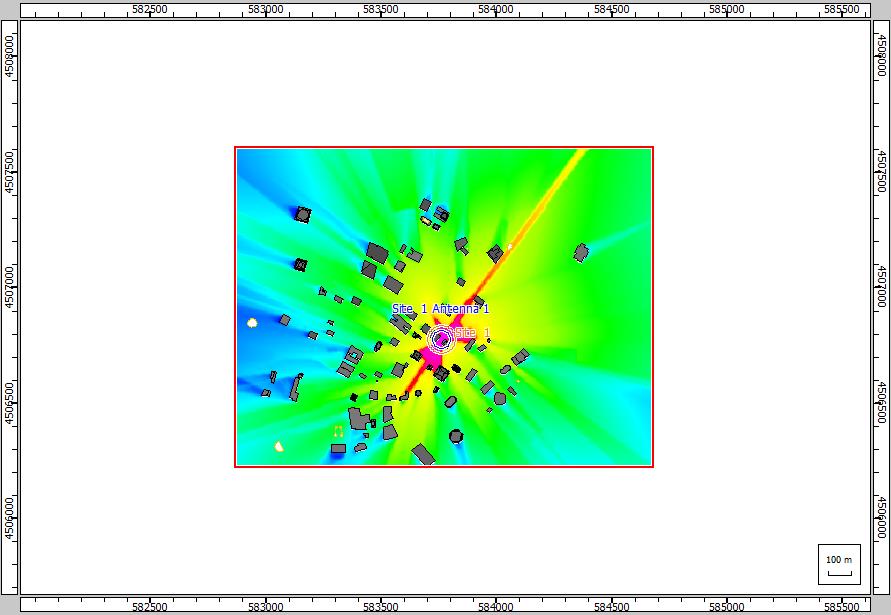}}
	\caption{Top view, 3D view and colored map of the path loss for the coverage areas for Suburban, Urban, Dense-urban and high-rise building environments when UAV altitude is at 120 m.}
	\label{figur3-3}
\end{figure*}

To assess Millimeter Wave propagation in High-rise Urban, Dense Urban, Urban and Suburban environments, extensive measurements are conducted of 28 GHz and 73 GHz channels in (Victoria, Australia as in Figure~\ref{C1-figur:3}), (Paris, France as in Figure~\ref{C2-figur:3}), (Mumbai, India as in Figure~\ref{C3-figur:3}), and (New York, USA as in Figure~\ref{C4-figur:3}) respectively. The 28 GHz and 73 GHz are suitable candidates for early Millimeter Waves deployments. 
Previously, Local Multipoint Delivery Systems were aimed at the 28 GHz frequency band and now consider it an enticing opportunity for initial cellular deployments of the Millimeter Wave due to its relatively low frequency within the Millimeter Wave spectrum~\cite{akdeniz2014millimeter}. 
Moreover, the 73 GHz frequency band has an abundant spectrum that can be adapted for dense deployment and can accommodate further expansions if low-frequency bands are congested~\cite{akdeniz2014millimeter}.

To measure the channel characteristics at these frequencies, an uplink scenario for GTA communication system is considered, and WinProp software is used to conduct the experiments where the transmitter hight is 1.7 meter and the receiver (UAV) hight is 120 meters. As a function of the transmitter-receiver distance, the total path loss is calculated.
At each placement, the path loss is estimated as:
\begin{equation}
PL=P_t-P_r+G_t+G_r,
\end{equation}
where $P_t$ is the transmitter power, $P_r$ is the receiver power and $G_t$ and $G_r$ are the gains of the transmitter and receiver antennas, respectively. For this
experiment, $P_t$ = 40 dBm and $G_t$ = $G_r$ = 0 dB.

\begin{figure*}[!h]
	\centering
	\subfloat[Suburban at 28 GHz.]{\label{figur:1}\includegraphics[width=.33\textwidth]{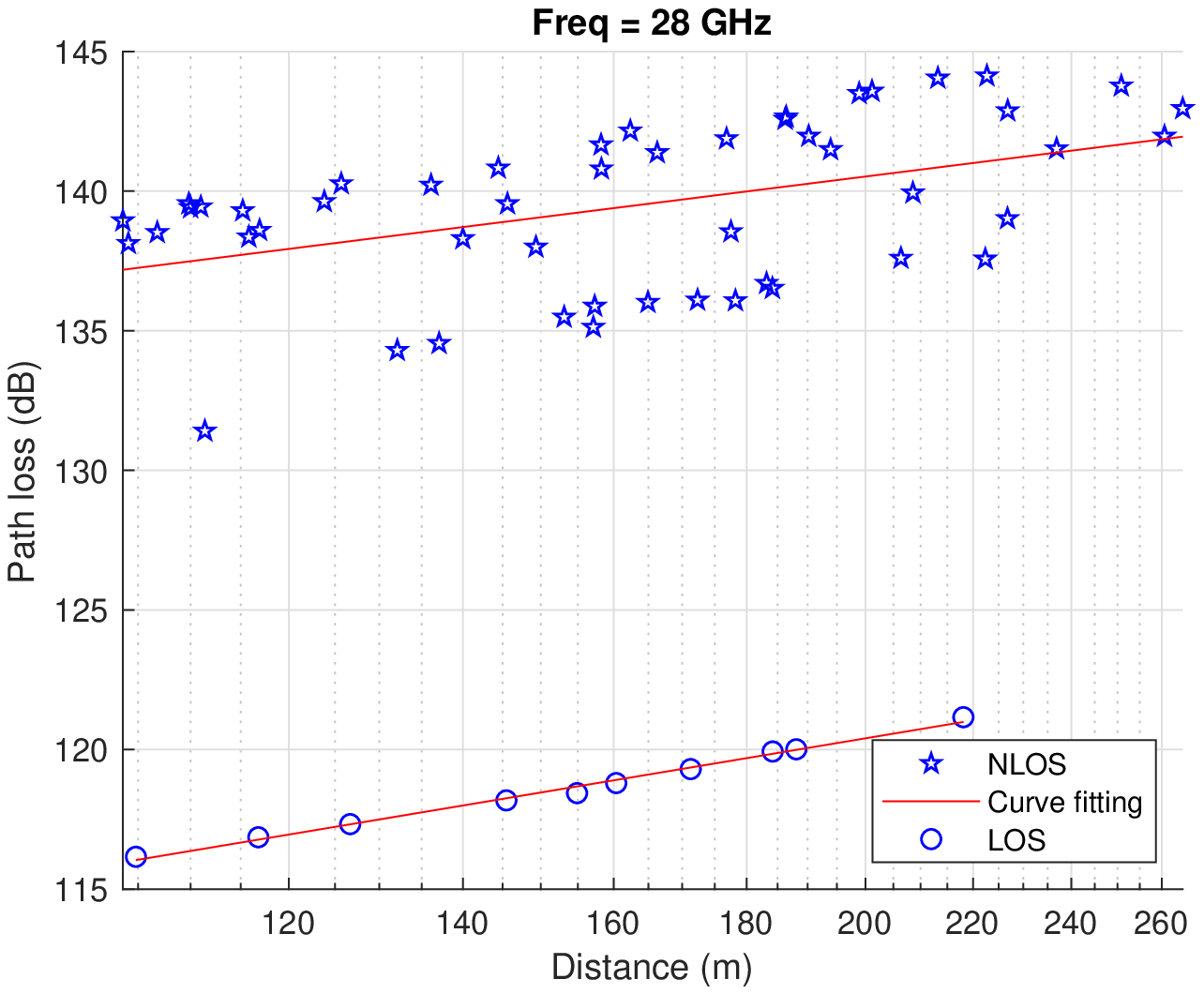}}
	\subfloat[Suburban at 73 GHz.]{\label{figur:2}\includegraphics[width=.33\textwidth]{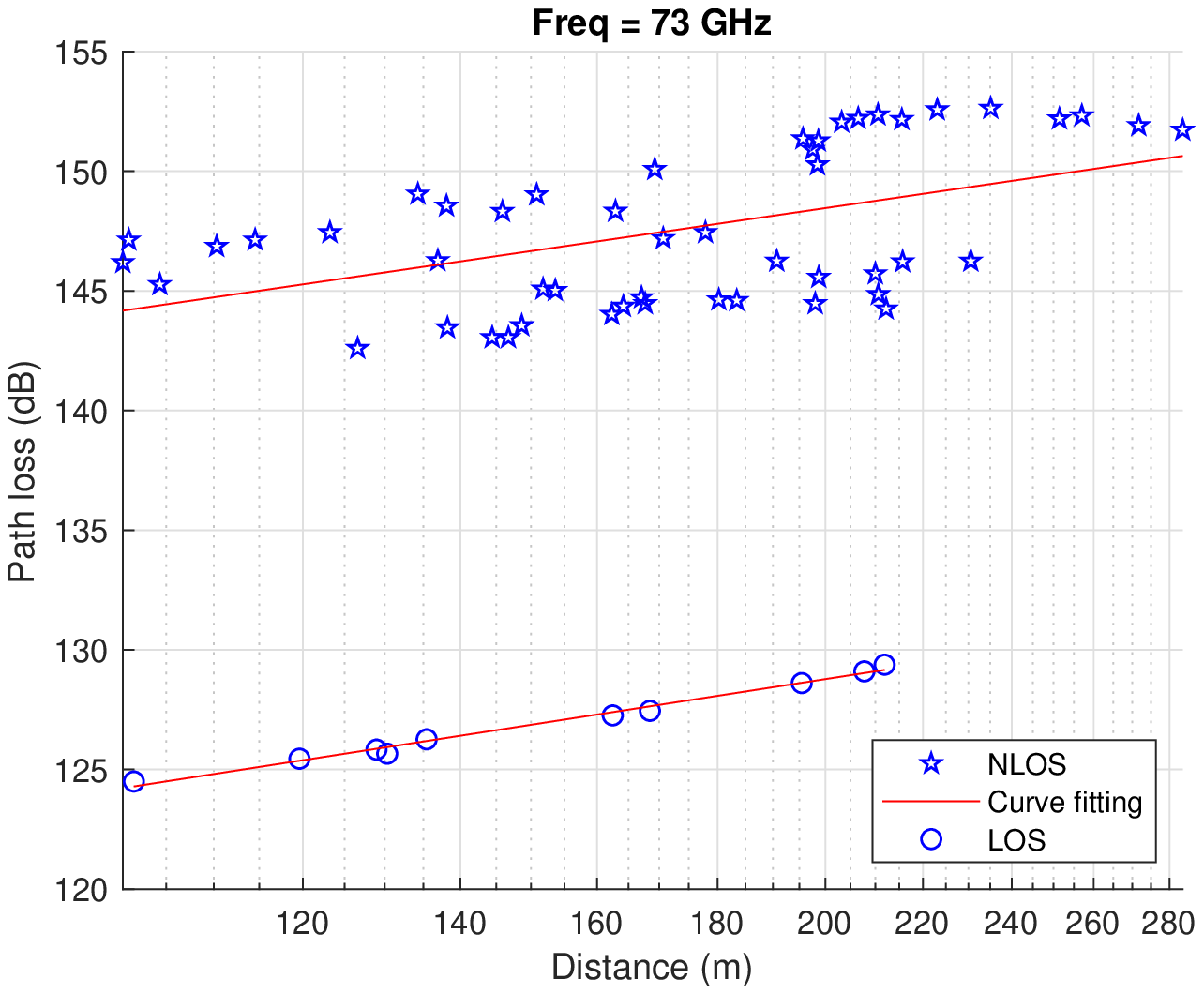}}
	\\
	\subfloat[Urban at 28 GHz.]{\label{figur:3}\includegraphics[width=.33\textwidth]{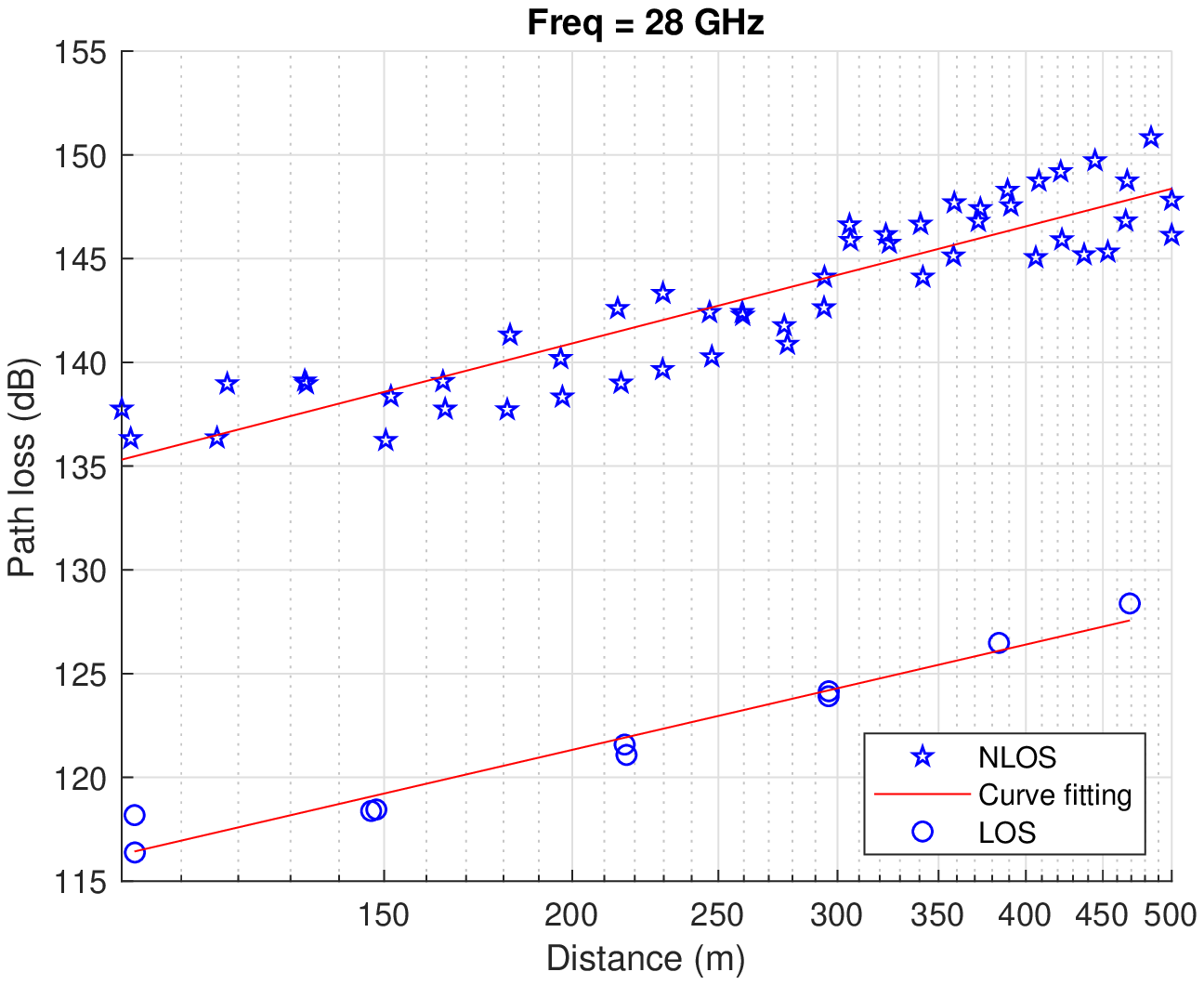}}
	\subfloat[Urban at 73 GHz.]{\label{figur:4}\includegraphics[width=.33\textwidth]{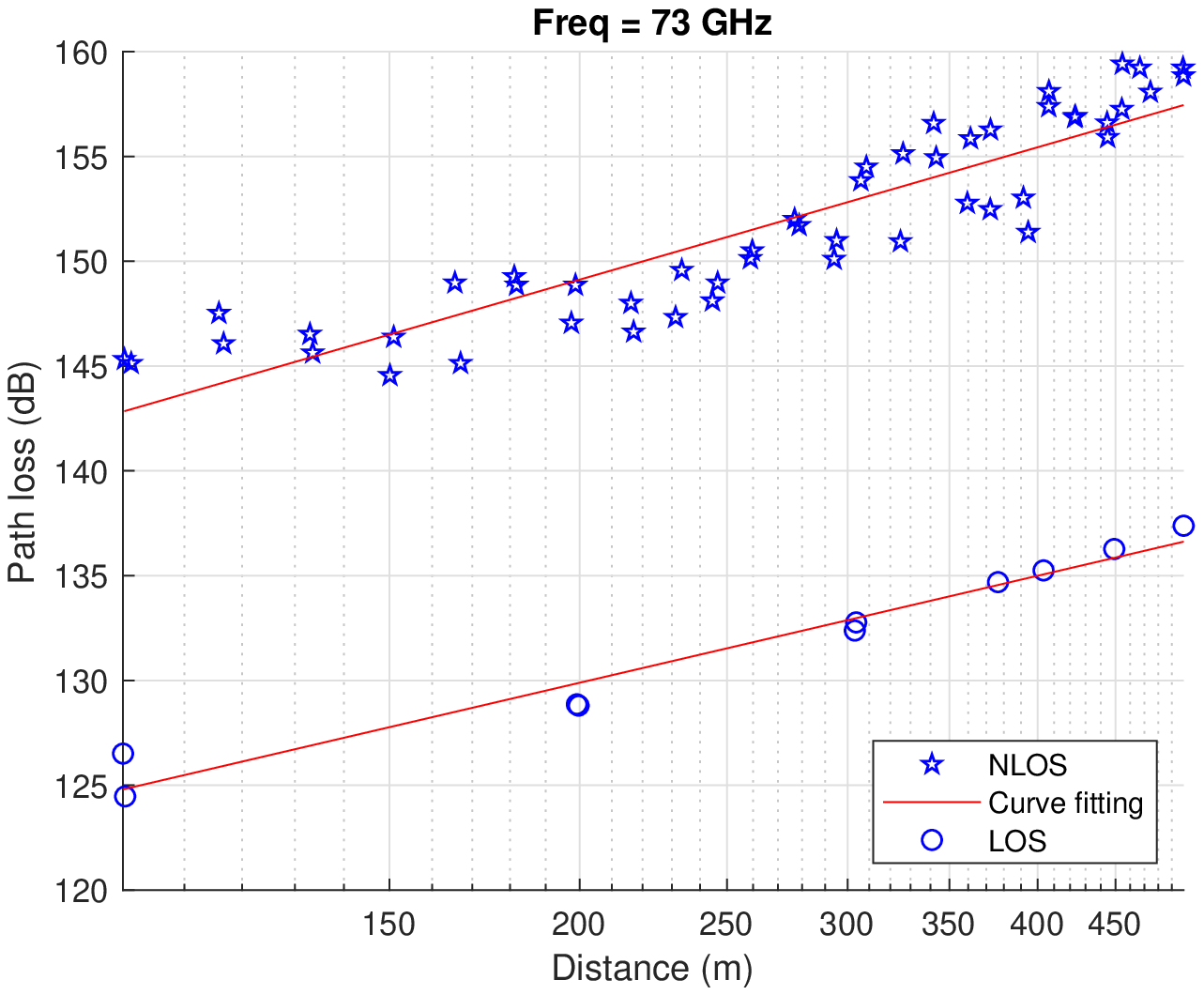}}
	\\
	\subfloat[Dense-urban at 28 GHz.]{\label{figur:5}\includegraphics[width=.33\textwidth]{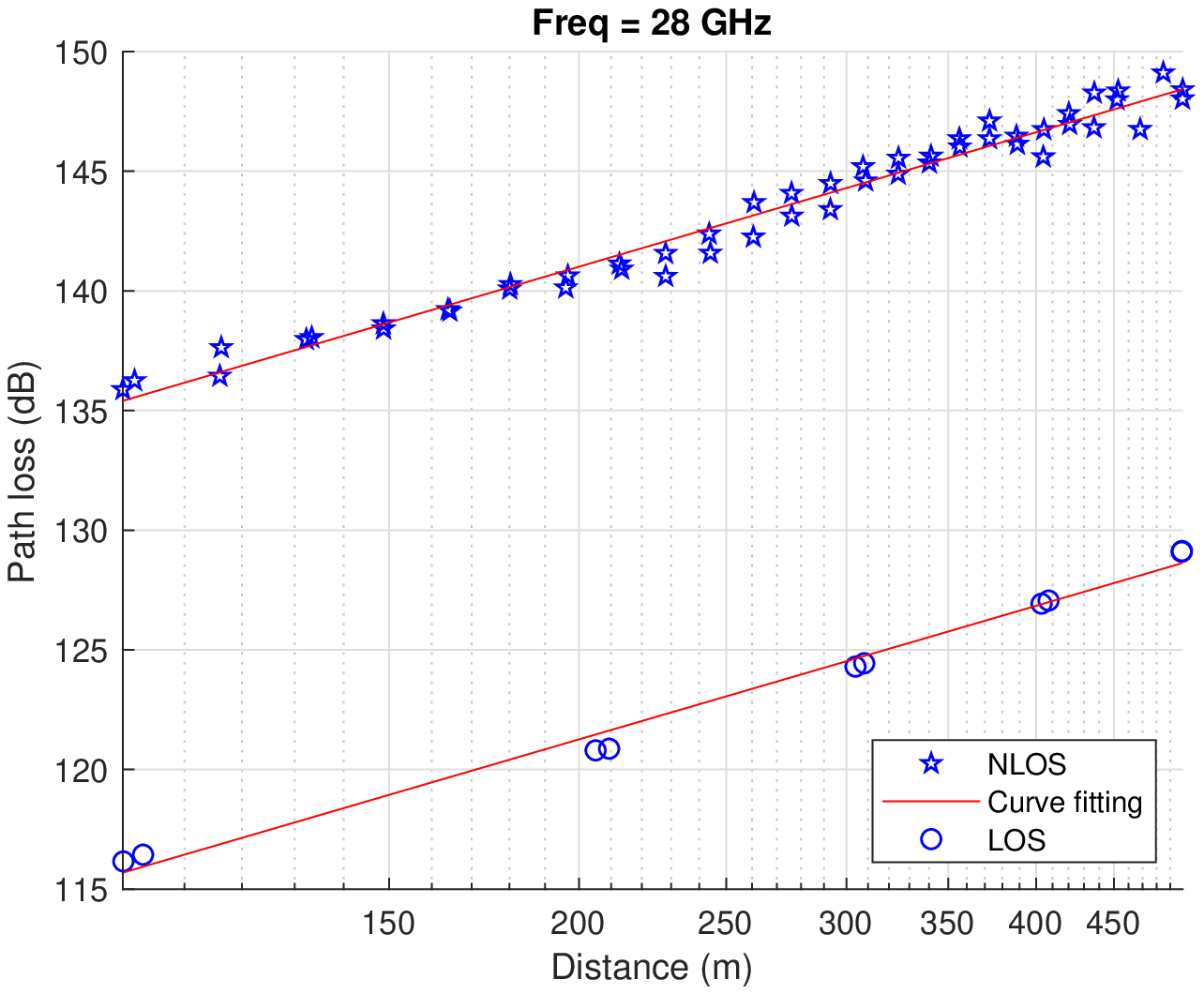}}
	\subfloat[Dense-urban at 73 GHz.]{\label{figur:6}\includegraphics[width=.33\textwidth]{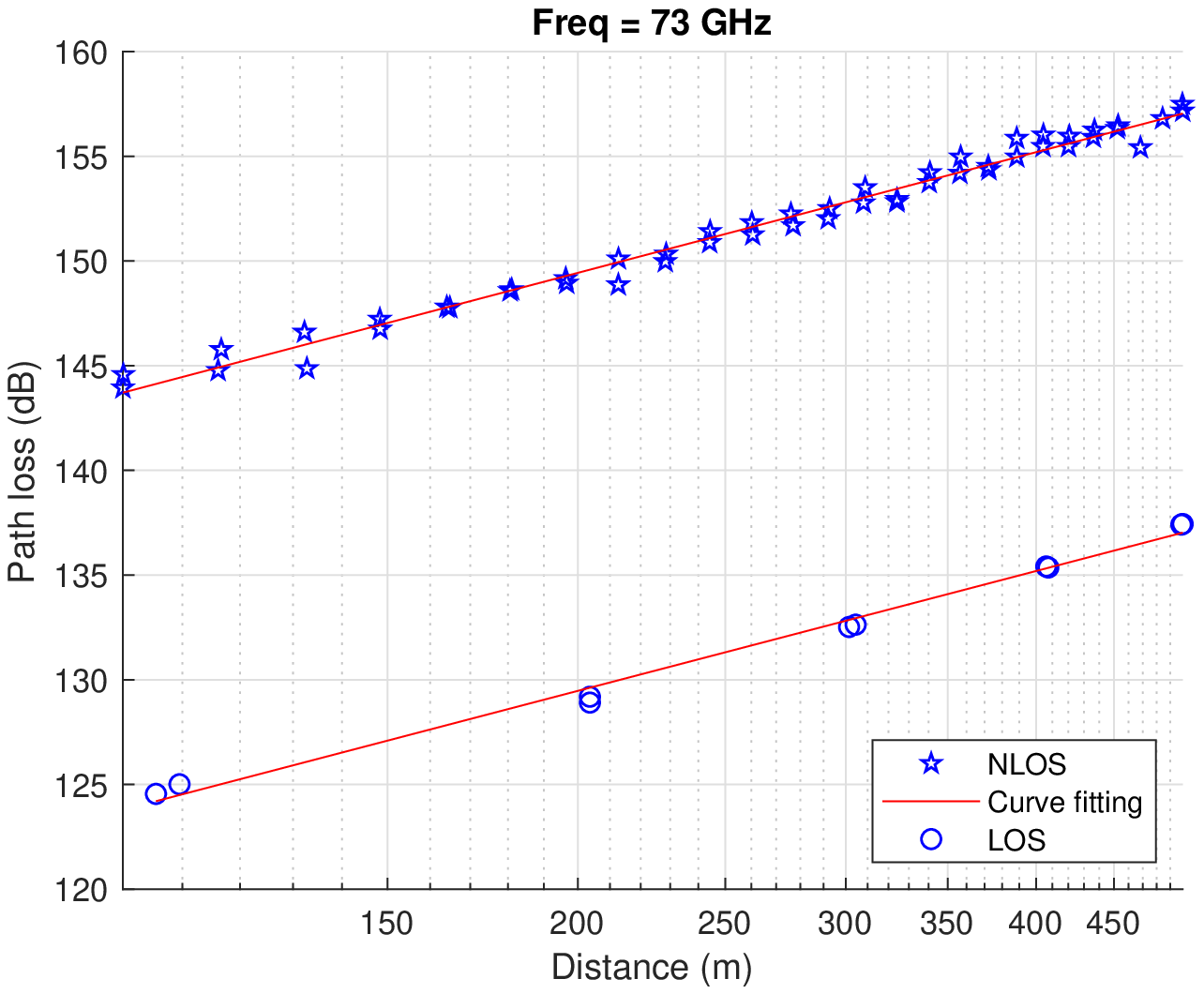}}
	\\
	\subfloat[High-rise buildings at 28 GHz.]{\label{figur:5}\includegraphics[width=.33\textwidth]{case3-DenseUrban-28ghz-curve-fitt.eps}}
	\subfloat[High-rise buildings at 73 GHz.]{\label{figur:6}\includegraphics[width=.33\textwidth]{case3-DenseUrban-73ghz-curve-fitt.eps}}
	\caption{Scatter plot with linear fit of the estimated path loss for different environments for LOS and NLOS links at 28/73 GHz. }
	\label{figur2-2}
\end{figure*}

As a function of the transmitter-receiver line of sight distance, a scatter plot of the path losses at different placements is shown in Figure~\ref{figur2-2}. Each placement is manually labeled in this experiment as either line of sight, where the transmitter is visible to the receiver, or non-line of sight, where the transmitter is obstructed. It is normal to fit the line of sight and non-line of sight path losses separately in traditional cellular path loss models~\cite{akdeniz2014millimeter}.


In Figure~\ref{figur2-2}, we plot a fit using a standard linear model for the line of sight and non-line of sight points,
\begin{equation}
PL(d) [dB]=\alpha+\beta10log_{10}(d)+\zeta,~~~\zeta\sim \mathcal{N}(0,\sigma^2),
\end{equation}
where $d$ is the 3D distance between the ground user and the UAV (meters), the least square fits of floating intercept and slope over the calculated distances (200 to 500 m) are $\alpha$ and $\beta$, and $\sigma^2$ is the variance of lognormal shadowing. The values of $\sigma^2$, $\alpha$ and $\beta$ are shown in Table~\ref{table1-1} for 28 GHz, and Table~\ref{table2-2} for 73GHz.

In Tables ~\ref{table1-1}~and~\ref{table2-2}, we can notice that the path loss increases as frequency and distance increase. For example, the values of $\alpha$ and $\beta$ parameters are 97.81 and 1.87 for the dense environment when the frequency is 28 GHz, while the values of $\alpha$ and $\beta$ parameters are 100.83 and 2.09 for the dense environment when the frequency is 73 GHz. Also, we can notice in Figure~\ref{figur3-3} that the path increases as distance increase. Moreover, we can notice that the line of sight parameters are close together, this is because the transmitter is visible to the receiver. Also, we can notice that the parameter differences for urban and dense urban environments are small, this because we do not take into account the density of users when we derive the Ground-to-Air path loss model, where the human body is considered as a blocker in Millimeter Wave networks.

To account for the blockage of the human body, we utilize
the probability of line of sight, $P_L$, for a wireless device $i$ from~\cite{gapeyenko2018effects} as:
\begin{equation}
P_L(r_i,h_D)=exp{(-\lambda g_B ({r_i(h_B-h_R)}/{(h_D-h_R)}))},
\end{equation}
where $r_i$ is 2D distance between a ground user and the UAV, $h_D$ is the hight of the UAV, $\lambda$ is the density of human blockers, $g_B$ is the diameter of human blockers, $h_B$ is the hight of the human blocker, $h_R$ is the hight of the wireless device (transmitter).

The average path loss between a ground wireless device $i$ and the UAV is given by:
\begin{equation}
PL_{a,i}=P_L(r_i, h_D)PL_{L,i}+[1-P_L(r_i, h_D)]PL_{N,i},
\end{equation}
where $PL_{L,i}$ and $PL_{N,i}$ are the path losses for line of sight and non-line of sigt links.

\begin{table}[!h]
	\caption{Environment Parameters for GTA Path loss Model -  Frequency 28 GHz.}
	\centering
	\resizebox{\columnwidth}{!}{
		\begin{tabular}{cccccc}
			\hline
			\multirow{2}{*}{Fitting Parameters} & \multirow{2}{*}{Link Type } & Sub-Urban. & Urban & Dense-Urban. & High-rise Building\\
			
			&                   & Victoria, Australia   & Paris,France  &Mumbai, India   &
			New York, USA  \\
			
			\hline
			$\alpha$  & \multirow{3}{*}{NLOS} &  113.63 &	97.81	& 98.05 & 66.25 \\
			$\beta$  &                   & 1.16&	1.87&	1.86&	3.30  \\
			$\zeta\sim\mathcal{N}(0,\sigma^2) $   &                   &  2.58&	1.69&	0.59&	4.48  \\
			\hline
			$\alpha$   & \multirow{3}{*}{LOS} &  84.64&	82.54&	78.58&	88.76  \\
			$\beta$  &                   &  1.55&	1.68&	1.85&	1.68 \\
			$\zeta\sim\mathcal{N}(0,\sigma^2) $    &                   &  0.12&	0.79&	0.49&	2.47 \\
			\hline
		\end{tabular}
	}
	\label{table1-1}
\end{table}

\begin{table}[!h]
	\caption{Environment Parameters for GTA Path loss Model - Frequency 73 GHz.}
	\centering
	\resizebox{\columnwidth}{!}{
		\begin{tabular}{cccccc}
			\hline
			\multirow{2}{*}{Fitting Parameters} & \multirow{2}{*}{Link Type } & Sub-Urban. & Urban & Dense-Urban. & High-rise Building\\
			
			&                   & Victoria, Australia   & Paris,France  &Mumbai, India   &
			New York, USA  \\
			
			\hline
			$\alpha$  & \multirow{3}{*}{NLOS} &  115.40&	100.83&	105.37&	102.10 \\
			$\beta$  &                   &  1.43&	2.09&	1.91&	2.22 \\
			$\zeta\sim\mathcal{N}(0,\sigma^2) $     &                   &  2.74&	1.90&	0.46&	6.61  \\
			\hline
			$\alpha$   & \multirow{3}{*}{LOS} &  93.63&	90.86&	85.71&	85.49  \\
			$\beta$  &                   &  1.52&	1.69&	1.90&	1.92 \\
			$\zeta\sim\mathcal{N}(0,\sigma^2) $   &                   &  0.16&	0.84&	0.42&	0.57 \\
			\hline
		\end{tabular}
		\label{table2-2}
	}
\end{table}


\section{Conclusions}
\label{con}
In this research work, a mathematical propagation model for estimating the Ground-to-Air path loss between a wireless system and a low altitude platform was provided utilizing millimeter-wave frequency bands. This model was developed based on simulation experiments in four different environments: high-rise buildings, dense-urban, urban and suburban in New York, USA, Mumbai, India, Paris, France, and Victoria, Australia, respectively, at 28 GHz and 73 GHz. 
In this model, a scatter plot of the path losses was presented at different placements as a function of the transmitter-receiver distance for the line of sight and non-line of sight.
Each placement was manually labeled as either line of sight, where the transmitter is visible to the receiver, or non-line of sight, where the transmitter is obstructed. The proposed model of path loss can help to formulate several significant problems for academic researchers. 
In our future works, the human blocker factor will be considered in the path loss model. Furthermore,  different mmWave frequency bands will be studied. 
Moreover, we will conduct real experiments and compare their results with simulation results.  
	
	\bibliographystyle{IEEEtran}
	\bibliography{01-paper_preprint}
	
	
\end{document}